\documentclass[twocolumn]{aastex631}

\usepackage{amsmath}
\begin{document}
\title{Quasi-periodic $\gamma$-ray modulations in the blazars PKS 2155-83 and\\ PKS 2255-282}
\correspondingauthor{M. A. Hashad}
\email{Mohamed.Hashad@bue.edu.eg}
\author{M. A. Hashad}
\affiliation{Basic Sciences Department, Modern Academy for Engineering and Technology, Maadi, 11585, Cairo, Egypt}
\affiliation{Centre for Theoretical Physics, The British University in Egypt, Sherouk City, 11837, Cairo, Egypt}
\author{ Amr A. EL-Zant}
\affiliation{Centre for Theoretical Physics, The British University in Egypt, Sherouk City, 11837, Cairo, Egypt}
\author{Y. Abdou}
\author{H. M. Badran}
\affiliation{Physics Department, Faculty of Science, Tanta University, Tanta, 31527, Gharbia, Egypt}
\begin{abstract}
\noindent While there has been an increase in interest in the possibility of quasi-periodic oscillations (QPOs) in blazars, the search has hitherto been restricted to sources with well-sampled light curves. Objects with light curves that include gaps have been, to our knowledge, overlooked. Here, we study two such curves, which have the interesting feature of pertaining to relatively high redshift blazars --- FSRQs, PKS 2155-83 and PKS 2255-282 --- observed by \textit{Fermi}-LAT. Their redshifts border the ‘cosmic noon’ era of galaxy formation and merging, and  their light curves exhibit a distinctive pattern of repetitive high and low (gap dominant) states for $15.6$ years. To accommodate for the gaps in the curves, data is integrated over extended time intervals of 1 month and 2 months. The resulting curves were also examined using methods suitable for sparsely sampled data. This investigation of PKS 2155-83 and PKS 2255-282 suggests QPOs with periods of $4.69\pm0.79$ yr ($3\sigma$) and $6.82\pm2.25$ yr ($2.8\sigma$), respectively. The PDFs of the blazars’ fluxes, along with the correlation between their flux and spectral index, were also analyzed. Given the epochs the objects are observed, the plausibility of a binary black hole scenario as an origin of the apparent periodicity was examined. We estimated the prospective parameters of such a system using a simple geometric model. The total masses were estimated, and found to be consistent, in principle, with independent (dynamical) measurements of the central black hole masses in the two host galaxies.
\end{abstract}
\section{Introduction} \label{sec:intro}
Collimated plasma jets are launched from active galactic nuclei (AGN) born of spinning black holes and strongly magnetized accretion disks. Blazars are a class of AGN whose relativistic jets nearly point to the line of sight connecting it to the Earth, and are dominated by non-thermal emission \citep{Blandford2019, Madejski2016, Urry1995}. Owing to the jet’s alignment and its narrow opening, Doppler enhancement is expected of the blazar emission and, subsequently, contract the timescales of its variability \citep{Yan2018}.

Erratic variability in blazars' emission has been observed over almost all the electromagnetic spectrum and over a wide range of timescales \citep{Abhir2021, Bhatta2020, Liao2014, Błażejowski2005}. In particular,  virtue of the continuous all-sky monitoring of the Fermi Large Area Telescope (\textit{Fermi}-LAT; \citealt{Atwood2009}), launched in 2008, one can seek out and check for quasi-periodicity in $\gamma$-ray blazars with periods up to a few years.
Numerous recent results have indeed found evidence for the existence of such quasi-periodic variability in blazars \citep{Ren_2023, Hashad2023, Zhang2020, Peñil2020, Benkhali2020, Tavani2018,Sandrinelli2018, Prokhorov2017}. 

After six years of \textit{Fermi}-LAT data gathering, the first evidence of a significant QPO ($\sim 3\sigma$) in $\gamma$-ray LC has been reported for PG 1553+113 with a period of 2.18 yr \citep{Ackermann2015}. One remarkable case is PKS 2247-131, which manifests a short periodicity ($\sim$ 34.5 days) in its $\gamma$-ray LC from November 2016 to June 2017 with six cycles at high significance ($5.2\sigma$) \citep{Zhou2018}. This QPO has been interpreted in terms of a helical structure in the jet. Using 12 years of \textit{Fermi}-LAT data, \cite{peñil2022evidence} have examined the $\gamma$-ray LCs of the most promising 24 periodicities reported in the literature. Five blazars with $\gamma$-ray QPOs observed with significance $\gtrsim 3\sigma$ have been found, one FSRQ, PKS 0454-234, and four BL Lacs, OJ 014, PG 1553+113, S5 0716+714, and PKS 2155-304. Observing such periodic signals could provide insight into blazars' nature and black hole (BH)-jet systems.

Indeed, the mechanism producing possible $\gamma$-ray QPO in blazars is not  entirely understood. Scenarios like jet precession, pulsational accretion flow instabilities, and the existence of binary black hole systems have been proposed \citep[e.g., ][]{Caproni2017, Sobacchi2016, Ackermann2015}. The broadness of the possibilities is reflected in the fact that, within the variety of models of QPOs in blazars, the QPOs can  originate from intrinsic as well as apparent origins; in the intrinsic scenario, QPO is assumed to exist in the relativistic jet’s co-moving frame, while an apparent origin refers to a periodically changing viewing angle and associated Doppler factor, which in turn boosts the observed flux periodically.

The reported searches for quasi-periodicity usually focus on objects with LCs that have significant detections in most bins. Typically, the minimum detection significance of the bins is set to be $\geq3\sigma$. As a result, important attributes in LCs with many upper limits or gaps are generally neglected. 

In this work, we introduce the results of a QPO search in two moderate redshifts blazars: flat-spectrum radio quasars (FSRQs), PKS 2155-83 ($z=1.865$), and PKS 2255-282 ($z=0.926$). Both exhibit the distinguishing behavior of repetitive high states and extremely faint epochs. These faint epochs are considered low (quiescent) states where upper limits and gaps frequently exist, particularly for time-binning of narrow time intervals. Here, we examine longer binning intervals and use methods that are suitable for time series with gaps, in order to attempt to reveal possible QPO signals.  
The redshifts, bordering the ‘cosmic noon’ era of galaxy formation and merging (in some definitions, 
PKS 2155-83 being well within that epoch), may hint at the apparent QPO being suitable candidates to be associated with supermassive black hole (SMBH) merger events \citep{Mezcua_2024-CosmicNoon, Yang_2020, Volonteri_2015, Madau_2014StarFormation, Komossa_2015, Begelman1980}. 

Our search is based on \textit{Fermi}-LAT data over about 15.6 years. A detailed analysis of \textit{Fermi}-LAT data is presented in Section \ref{Data Analysis and Results}. The resulting LCs are then examined for possible quasi-periodicities, using various methods, in Section \ref{Search for QPOs}. Section \ref{Summary and Discussion} summarizes the results and discusses the findings.

\section{Data Analysis} 
\label{Data Analysis and Results}

\subsection{The \textit{Fermi}-LAT Light Curves} 
\label{LAT Light Curves}

The Fermi Gamma-ray Space Telescope is a space mission with two scientific instruments: the LAT and the Gamma-ray Burst Monitor (GBM). The \textit{Fermi}-LAT high-energy $\gamma$-ray telescope covers the energy range from about 20 MeV up to 1 TeV and owns a large effective area ($\sim8000$ $\mathrm{cm^2}$ at 1 GeV), with $\sim 2.4$ sr field of view and a point-spread function (PSF) of $< 0.8^{\circ}$ above 1 GeV \citep{Atwood2009}. LAT scans the whole celestial sphere every 3 hours. The instrument took off on 11 June 2008, and is still in operation. In the course of its 15-year operation, it detected high-energy gamma rays from assorted classes of objects with the most severe conditions, including but not limited to gamma-ray blazars.

\begin{figure}[htp]
\centering
\includegraphics[width=1\columnwidth]{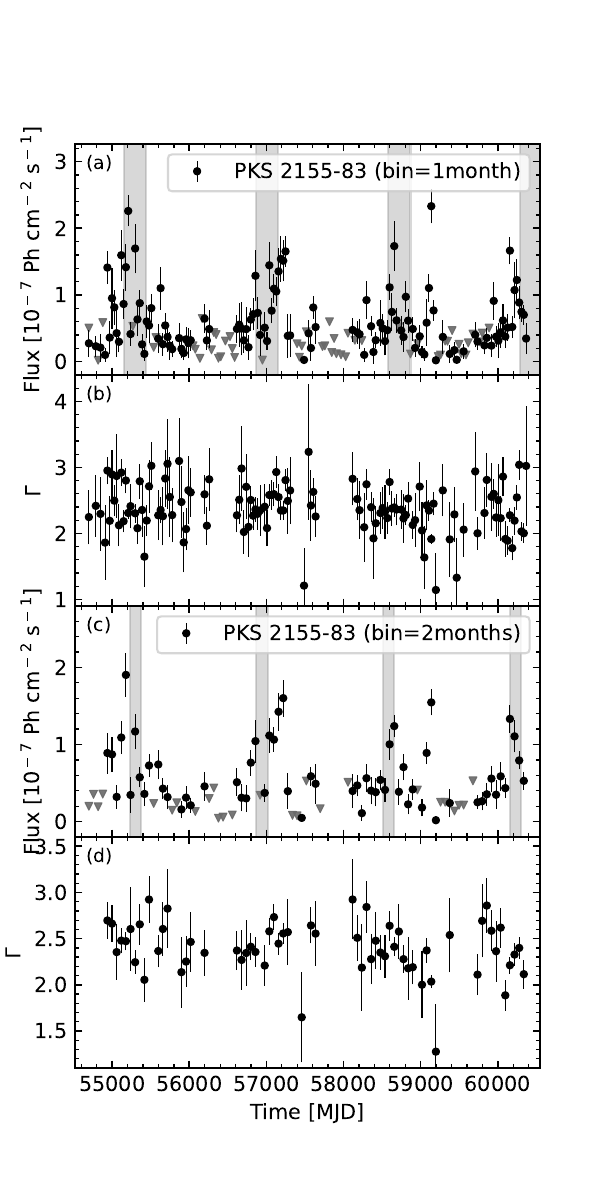}
\caption{PKS 2155-83 LCs in the energy range from 100 MeV to 500 GeV between 4 August 2008, and 29 February 2024. Showing (a) the one month (1-m) binned LC, (b) the photon spectral index of (a), (c) two months (2-m) binned LC, and (d) the photon spectral index of (c). In panels (a) and (c), filled black points denote significant detections, with $TS>4$ for the 1-m binned LC and $TS>9$ for the 2-m binned LC. Downward gray arrows denote $95\%$ confidence level upper limits. The gray vertical columns approximately delineate periods of high states, inferred from the generalized Lomb Scargle periodogram. The periodic signals' uncertainty is indicated by the width of the gray columns. The photon index is plotted only for the detected bins. In all panels, vertical error bars are $1\sigma$ error.}
\label{figLC1}
\end{figure}

We consider here the LCs of two blazars detected by the Fermi observatory, searching within them for quasi-periodic signals. The distant FSRQ PKS 2155-83 is at $RA=330\mathrm{h} 36\mathrm{m} 00.00\mathrm{s}$ and $Dec=83\mathrm{d} 35\mathrm{m} 60.0\mathrm{s}$, J2000 \citep{Fabricius2021, Mauch2003}. It was observed by the \textit{Fermi}-LAT in a high state on 5 January 2010, with a $\gamma$-ray flux $F(E > 100\ \mathrm{MeV})$ of $(1.4 \pm 0.3)\times 10^{-6}$ photons $\mathrm{cm^{-2}}$ $\mathrm{s^{-1}}$, which is more than an order of magnitude larger than the average flux during the first 11 months of observations \citep{Wallace2010}. According to the \textit{Fermi}-LAT 4th source catalog \citep[4FGL-DR3; ][]{Abdollahi_2022_4FGL-DR3}, PKS 2155-83 (4FGL J2201.5-8339) has an average detection significance of $41.88\sigma $ with a predicted photon number of 4247.39 and flux fractional variability of $67.86\pm 14.84 \%$. During the 15.6 years of \textit{Fermi}-LAT observations, the object has had eminent behavior, where it seems to release four high states in 2010, 2014, 2019, and 2023 throughout its low state, as shown in the LCs in Fig. \ref{figLC1}. This behavior may underline a featured origin, as the repeated high states look periodic.

\begin{figure}[htp]
\centering
\includegraphics[width=1\columnwidth]{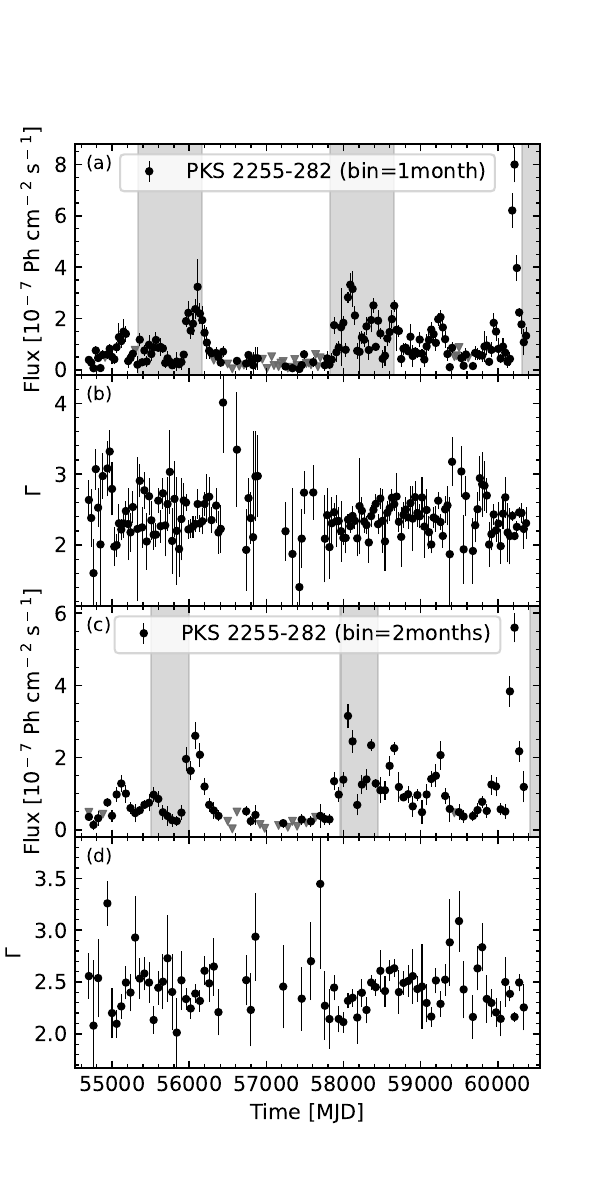} 
\caption{The same as Figure \ref{figLC1}, but for the object PKS 2255-282.}
\label{figLC2}
\end{figure}

Similarly, another distant FSRQ, PKS 2255-282 --- with $RA=344\mathrm{h} 30\mathrm{m} 00.00\mathrm{s}$ and $Dec=-27\mathrm{d} 53\mathrm{m} 60.0\mathrm{s}$, J2000 \citep{Jones2009} --- shows three high states (2009-2013, 2017-2021, and 2023-up to the end of data) throughout its low state, as shown in Fig. \ref{figLC2}. An average detection significance of $66.57\sigma $, a predicted photon number of 7383.77, and flux fractional variability of $84.67\pm 18.17\%$, were reported from the source in Fermi's fourth catalog \citep{Abdollahi_2022_4FGL-DR3}. Radio measurements at 15 GHz \citep{Lister2009} and 22 and 43 GHz \citep{Charlot2010} suggested a compact object with a core-dominated structure. The object was in a high state on 26 February 2012, with a $\gamma$-ray average daily flux above 100 MeV of ($1.0\pm 0.3)\times{10}^{-6}\ \mathrm{photons\ }{\mathrm{cm}}^{\mathrm{-}\mathrm{2}}~{\mathrm{s}}^{\mathrm{-}\mathrm{1}}$. This exemplifies a boost factor of $\mathrm{\sim}$ 14 above its average flux in Fermi's second catalog \citep{Dutka2012}. It was the first significant \textit{Fermi}-LAT source to be detected at such a high value of flux, although a $\gamma$-ray flare was detected earlier from the object by EGRET in  December 1997 \citep{Macomb1999}. {The EGRET outburst lasted from 30  December 1997 till 12 January 1998, with possible weak variability of a short timescale of several days \citep{Tornikoski_1999}. The total flux above 100 MeV was ($1.6\pm 0.3)\times{10}^{-6}\ \mathrm{photons\ }{\mathrm{cm}}^{\mathrm{-}\mathrm{2}}~{\mathrm{s}}^{\mathrm{-}\mathrm{1}}$ with a peak flux of ($4.8\pm 1.1)\times{10}^{-6}\ \mathrm{photons\ }{\mathrm{cm}}^{\mathrm{-}\mathrm{2}}~{\mathrm{s}}^{\mathrm{-}\mathrm{1}}$. This was higher than the quiescent upper limits to emission, based upon previous EGRET observations, by a factor of 20, placing PKS 2255-282 among EGRET’s brightest blazars. Before this outburst, PKS 2255-282 had been in the field of view of EGRET several times but was not detected in $\gamma$-rays, see Fig. 2 in \cite{Macomb1999}. In November 1997, EGRET recorded a flux above 100 MeV of ($4.7\pm 2.3)\times{10}^{-7}\ \mathrm{photons\ }{\mathrm{cm}}^{\mathrm{-}\mathrm{2}}~{\mathrm{s}}^{\mathrm{-}\mathrm{1}}$ (3$\sigma$ detection), which is about 10 times smaller than the flux from the outburst of  January 1998. The count rate dropped by roughly a factor of 3 by the end of the outburst period \citep{Tornikoski_1999}. As EGRET was in reduced field mode during the outburst period, the 9.2$\sigma$ detection only corresponds to $51\pm9$ source counts. With such sparse data, it was difficult to locate $\gamma$-ray variability. However, many photons were clustered around two separate times, from  2.2 to 3.1 January 1998 and from 9.1 to 10.5 January 1998. It is noteworthy that a prolonged quiescent state with reported upper limits preceded the outburst observed by EGRET and this was, to some extent, repeatedly observed later on by \textit{Fermi}-LAT.}

For the sake of performing the periodicity search for both objects PKS 2155-83 and PKS 2255-282, we proceed as follows: We generated LCs with one month (1-m) and two months (2-m) of time binning, encompassing the approximately 15.6 years (MJD: 54683--60369) of LAT data in the energy range 100 MeV--500 GeV (Figures \ref{figLC1} and \ref{figLC2}). A 1-m binned LC with detected bins contingent on test statistics $>4$ ($TS=2\mathrm{log}(L/L_{0})$, where $L$ and $L_{0}$ are the maximum likelihood of the models with and without a source at the target position) was constructed to allow inspection of the flux variation in relatively moderate details. The choice of 2-m binning with $TS>9$ is motivated by keeping the time intervals long enough to diminish the missing values, reduce fluctuations, and provide better statistics of the underlying variations within the data on the cost of losing the time resolution. The detection ratios (the number of detections in the LC to the total number of bins) in 1-m and 2-m are comparable and estimated to be about $64\%$ and $81\%$ for PKS 2155-83 and PKS 2255-282, respectively.
The LCs were then reduced with the maximum likelihood method using Fermitools version 2.2.0\footnote{\url{http://fermi.gsfc.nasa.gov/ssc/data/analysis/software}} by the implementation of the Python package fermipy\footnote{\url{http://fermipy.readthedocs.io}} \citep[Version: 1.2.2; ][]{Wood_2017}. The instrument response function P8R3\_SOURCE\_V3 was used with `SOURCE' class photons.

In this analysis, the photons within the region of interest (a $15^{\circ}\times 15^{\circ}$ square centered on the position of the source of interest) were selected. Photons were then modeled by accounting for the point sources in the 4FGL catalog that positioned around the source of interest (up to $20^{\circ}$). Moreover, the background emission was also modeled with including a galactic component (the Milky Way's diffuse $\gamma$-ray emission; $gll\_iem\_v07$.fits file) and an extragalactic one (the isotropic $\gamma$-ray emission from celestial and residual charged-particle backgrounds; $iso\_P8R3\_SOURCE\_V3\_v1$.txt file). A cut on the zenith angle larger than $90^{\circ}$ was imposed to exclude $\gamma$-ray augmentation from the earth limb. Additionally, we used the recommended data quality cuts (DATA QUAL$>0$)\&\&(LAT CONFIG$==1$) and removed time periods coinciding with gamma-ray bursts and solar flares detected by the LAT. A $0.1^{\circ}$ spatial binning and eight logarithmic energy bins per decade were adopted. The normalizations of all sources within $3^{\circ}$ away from the ROI center and the galactic and isotropic diffuse backgrounds, as well as the normalization and spectral index of the target source, were left free to vary in the likelihood analysis over the full time range of the observation. All other parameters were set at their catalog values. The routines gta.optimize() and gta.fit() were iteratively run till a good fit quality is achieved (fit$\_$quality = 3). To construct the LCs, we split the data for each source into 1-m and 2-m bins and conducted a full likelihood fit in each bin. This is done while utilizing the parameters' values obtained from the full time range analysis. The spectral parameters of the target except the scale parameter were left free during the fit. The normalizations of sources within $3^{\circ}$ from the center of the ROI along with the normalizations of diffuse components, were left to vary.

The observed $\gamma$-ray fluxes of both objects display striking variation (Figures \ref{figLC1} and \ref{figLC2}). The average flux of the 1-m binned LCs is; $F_{av}=\left(6.12\pm 4.64\right){\times 10}^{-8} \mathrm{~photons~}{\mathrm{cm}}^{\mathrm{-}\mathrm{2}}~{\mathrm{s}}^{\mathrm{-}\mathrm{1}} $~and~$\ F_{av}=(1.05\pm 1.02){\times 10}^{-7} \mathrm{\ photons\ }{\mathrm{cm}}^{\mathrm{-}\mathrm{2}}~{\mathrm{s}}^{\mathrm{-}\mathrm{1}}$ for PKS 2155-83 and PKS 2255-282, respectively and for the 2-m binned LCs is $F_{av}=(6.16\pm 0.53)\times {10}^{-8}\mathrm{\ photons\ }{\mathrm{cm}}^{\mathrm{-}\mathrm{2}}~{\mathrm{s}}^{\mathrm{-}\mathrm{1}}\ $ and $\ F_{av}=(1.04\pm 0.90)\times {10}^{-7}\mathrm{\ photons\ }{\mathrm{cm}}^{\mathrm{-}\mathrm{2}}~{\mathrm{s}}^{\mathrm{-}\mathrm{1}}$ for PKS 2155-83 and PKS 2255-282, respectively.

Photon spectral indices ($\mathrm{\Gamma }_{av}$) are estimated to be $2.39\pm 0.39$ (1-m) and $2.40\pm 0.29$ (2-m) for PKS 2155-83, and $2.41\pm 0.34$ (1-m) and $2.45\pm 0.26$ (2-m) for PKS 2255-282. Both sources show soft intrinsic $\gamma$-ray spectra like most FSRQs \citep{Madejski2016} for both 1-m and 2-m binned LCs. This probably explains the absence of $\gamma$-ray detection from these sources at higher energies. PKS 2155-83 (Pearson correlation coefficient $\rho_{1-m}=0.28$) and PKS 2255-282 ($\rho_{1-m}=0.01$) showed no flux-photon index correlation along the full time range of observation implying that both objects may favor the association with apparent geometrical effects.

\subsection{Flux Distribution} 
\label{Flux Distribution}

An essential feature of an astronomical source's variability is the distribution of the photon flux. The distinction between the two most common Gaussian and log-normal distributions can help characterize the inherent physical process causing the observed variability \citep{Shah2018, Rieger2019, Shah2020, Morris2019}. A Gaussian distribution reflects a linear random process, where the flux variation is to be indicated by the distribution width \citep{Sinha2018}. In this case, an additive statistical model is implied, with the linear summation of components taking part in building up the emission (e.g., shot-noise or a linear summation of many 
``mini-jets"). Recent findings, on the other hand, preferentially advocate the log-normal distribution for blazars at various wavelengths and timescales \citep{Romoli2018, Shah2018, Abramowski2010, Sinha2017, Wang2023}.

The log-normality of blazars' fluxes can be linked to the existence of a multiplicative process \citep{Rieger2019}. Accretion disk fluctuations could be a possible origin for such action, as in the case of X-ray binaries \citep{Lyubarskii1997, Arevalo_2006}. In this case, the accretion rate of mass varies as the result of an independent accretion disk's density fluctuations on a timescale that corresponds to the local viscous time scales. The fluctuations proliferate inward and provide a multiplicative process as they couple together in the innermost part of the disk \citep{Lyubarskii1997, King2004, Arevalo_2006}. If the instabilities in the accretion flow exhibit a quasi-periodic nature, the resulting QPOs will also propagate to the jet, and corresponding emission may be observed \citep{Rieger_2010}. By analogy, such origins for the log-normality may favor a binary black hole scenario for the possible quasi-periodic time signal. Log-normal flux distributions may also originate from cascade-related scenarios, such as magnetospheric inverse-Compton pair production cascades \citep{Levinson2011} or proton-induced synchrotron cascades \citep{Mannheim1993}. Further, the log-normal distribution could also be traced back to the acceleration process itself, for example, with linear Gaussian fluctuations in the particle acceleration rate inside the region of acceleration \citep{Sinha2018}. Finally, it should be mentioned that additive processes in specific scenarios can also lead to flux log-normality such as the overall flux (experienced Doppler boosting) from a large number of mini jets within a jet with random orientations \citep{Biteau2012}.

 For each object, we constructed a flux histogram for the 1-m binned LC chosen over 2-m binned data because it simply provides a statistically more significant fit of the distribution. The probability density function (PDF) of PKS 2155-83 and PKS 2255-282 were fitted by Gaussian G($\mathrm{\phi}$) and log-normal L($\mathrm{\phi}$) distributions given by
\begin{equation}
G\left(\varphi \mathrel{\left|\vphantom{\varphi \mu ,\sigma }\right.\kern-\nulldelimiterspace}\mu ,\sigma \right)=\frac{1}{\sqrt{2\pi }\sigma }{\mathrm{exp}\left(-\frac{{\mathrm{(}\mathrm{\varphi }-\mu )}^2}{2{\sigma }^2}\right)\ }
\end{equation}

and
\begin{equation}
L\left(\varphi \mathrel{\left|\vphantom{\varphi \mu ,\sigma }\right.\kern-\nulldelimiterspace}\mu ,\sigma \right)=\frac{1}{\sqrt{2\pi }\sigma \varphi }{\mathrm{exp}\left(-\frac{{{(\mathrm{log_{10}} \left(\varphi \right)}-\mu )}^2}{2{\sigma }^2}\right),\ }
\end{equation}
where $\mu $ and $\sigma $ are the mean and standard deviation, respectively.

For PKS 2155-83 PDF, the log-normal distribution ($r_{1-m}^2=0.94$) seems to be preferred over a Gaussian ($r_{1-m}^2=0.88$), which invokes a nonlinear, multiplicative process for the underlying variability rather than additive models. On the contrary, the PKS 2255-282 PDF has a comparable degree of fitness with both distributions; $r_{1-m}^2=0.99$ and $=0.95$ for Gaussian and log-normal distributions, respectively. The Gaussian and log-normal fit parameters, variance, probability, and W-Statistic of both sources are listed in Table \ref{tableFit parameters} for 1-m and 2-m cases. The normality test for 1-m and 2-m flux histograms for both objects is rejected.

\begin{table*}[t]
\centering
\caption{Fit parameters of PKS 2155-83 and PKS 2255-282 PDFs associated with the log-normal and Gaussian distributions, and flux normality tests.}
\label{tableFit parameters}
\begin{tabular}{lccccccccc} \hline\hline
Object Name & \multicolumn{1}{c}{Binning} & \multicolumn{3}{c}{Log-normal} & \multicolumn{3}{c}{Gaussian} & \multicolumn{2}{c}{Normality Test} \\
& & $\mu$* & $\sigma $* & $r^2$ & $\mu$* & $\sigma $* & $r^2$ & $P_{\text{value}}$ & W-Statistic \\ \hline
PKS 2155-83 & 1-m & 0.49& 0.75& 0.94& 0.39& 0.26& 0.88& $\mathrm{<}$0.001 & 0.86\\
& 2-m & 0.52& 0.69& 0.89& 0.41& 0.21& 0.78& $\mathrm{<}$0.001 & 0.90\\ \hline
PKS 2255-282 & 1-m & 0.80& 0.81& 0.99& 0.28& 0.94& 0.95& $\mathrm{<}$0.001 & 0.71\\
& 2-m & 0.82& 0.76& 0.97& 0.57& 0.68& 0.96& $\mathrm{<}$0.001 & 0.77\\ \hline
\end{tabular}
\begin{minipage}{12cm}
\centering
\small
*In units of $\times 10^{-7}$ photon cm${}^{-2}$ s${}^{-1}$.
\end{minipage}
\end{table*}

\section{Search for $\gamma$-ray Quasi-periodicity}
\label{Search for QPOs}

Several approaches have been used to inspect the quasi-periodic variability in blazars \citep{Wang2022}. In the present study, different methods were utilized to search for periodicity in the LC. These are described in the following sub-sections.

\subsection{Auto-correlation Function}

The auto-correlation function (ACF) is a reliable method suitable for detecting non-sinusoidal periodicities. It involves the correlation of the time series with itself, i.e., with the same series lagged by one or more time units. Constant auto-correlation is associated with a system remaining in the same state from one observation to the next; rapidly decaying ACF indicates a high degree of randomness in the time series, while periodicity in the ACF reflects corresponding periodicity in the data.

The ACF between observations ($f_{i} (t)$) separated by \textit{$\tau$} (= 0, 1, 2, 3,{\dots}, \textit{N}) time steps is given by
\begin{equation}
\begin{split}
&ACF\left(\tau \right)= \\ 
&\frac{\sum^{N-\tau }_{i=1}{\left\{f_i\left(t\right)-{\hat{f}}_{1\to N-\tau }\right\}\left\{f_{i+\tau }\left(t\right)-{\hat{f}}_{1+\tau \to N}\right\}}}{\left(N-\tau \right){\sigma }^2},
\end{split}
\end{equation}
where ${\hat{f}}_{1\to N-\tau }$ and ${\hat{f}}_{1+\tau \to N}$ are the means of the first (from the first to $N-\tau $ observations), and the last (from $1+\tau $ to $N$ observations) $N-\tau $ of the data points, respectively.

We used pyzdcf\footnote{\url{https://pypi.org/project/pyzdcf}}, a Python module that is utilized for robustly estimating cross-correlation functions of astronomical time-series data that are sparse and unevenly sampled~\citep{pyzdcf_Alexander_1997}. A Savitzky-Golay filter\footnote{\url{https://docs.scipy.org/doc/scipy/reference/generated/scipy.signal.savgol_filter.html}} was also applied to smooth the ACF, which effectively decreases the low-frequency fluctuations while preserving the overall tendency of the signal \citep{Press_1990}. The signal’s period is the median of a list of periods computed from the intervals between successive maxima and minima. The uncertainty is determined using the equation proposed by \cite{McQuillan_2013} as
\begin{equation}
{\sigma }_P=\frac{1.483\times MAD}{\sqrt{N-1}},
\end{equation}
where $MAD$ is the periods' median of the absolute deviations implied from the peaks list and $N$ is the number of peaks in the correlation. To determine the significance, we simulated ${10}^5$ LCs, using Emmanoulopoulos' method \citep{Emmanoulopoulos2013} as coded {in Python in \cite{Connolly2015}, that match both the power spectral density and probability density function of the object's LC. For each simulated LC, the ACF was applied, and the percentile was computed for each period to estimate the power confidence level.

The obtained ACFs are shown in Fig.~\ref{figACF}, {for 1-m binned LCs}. For the 1-m and 2-m binned LCs of PKS 2155-83, the estimated period is at $T=4.79\pm 0.35\ \mathrm{yr}\ (2.8\sigma)$ and at $T=4.68\pm 0.24\ \mathrm{yr}\ (1.8\sigma)$, respectively; and for the 1-m and 2-m binned LCs of PKS 2255-282 is at $T=6.53 \pm 1.34\ \mathrm{yr}\ (3.3\sigma)$ and at $T=6.27 \pm 0.16\ \mathrm{yr}\ (3\sigma)$, respectively.

\begin{figure}[ht]
\centering
{\includegraphics[width=1\columnwidth]{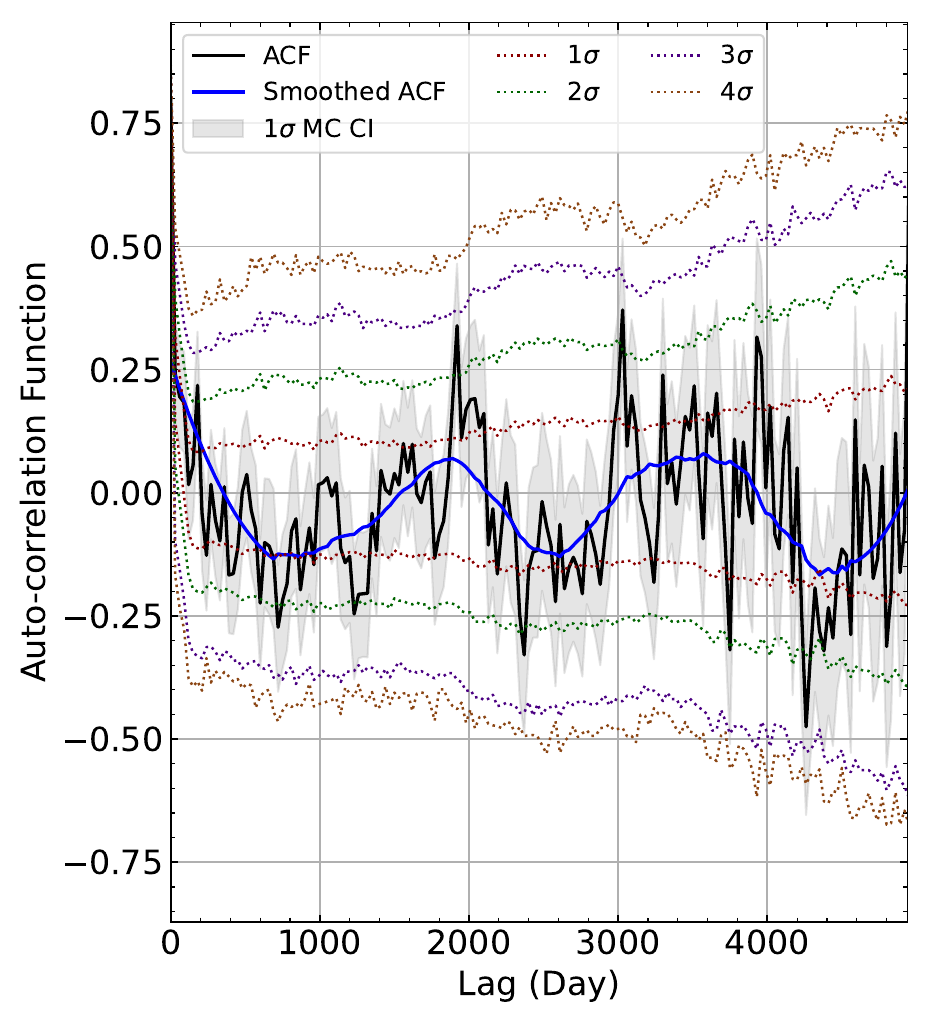}\label{figACF:subfigure1}}
{\includegraphics[width=1\columnwidth]{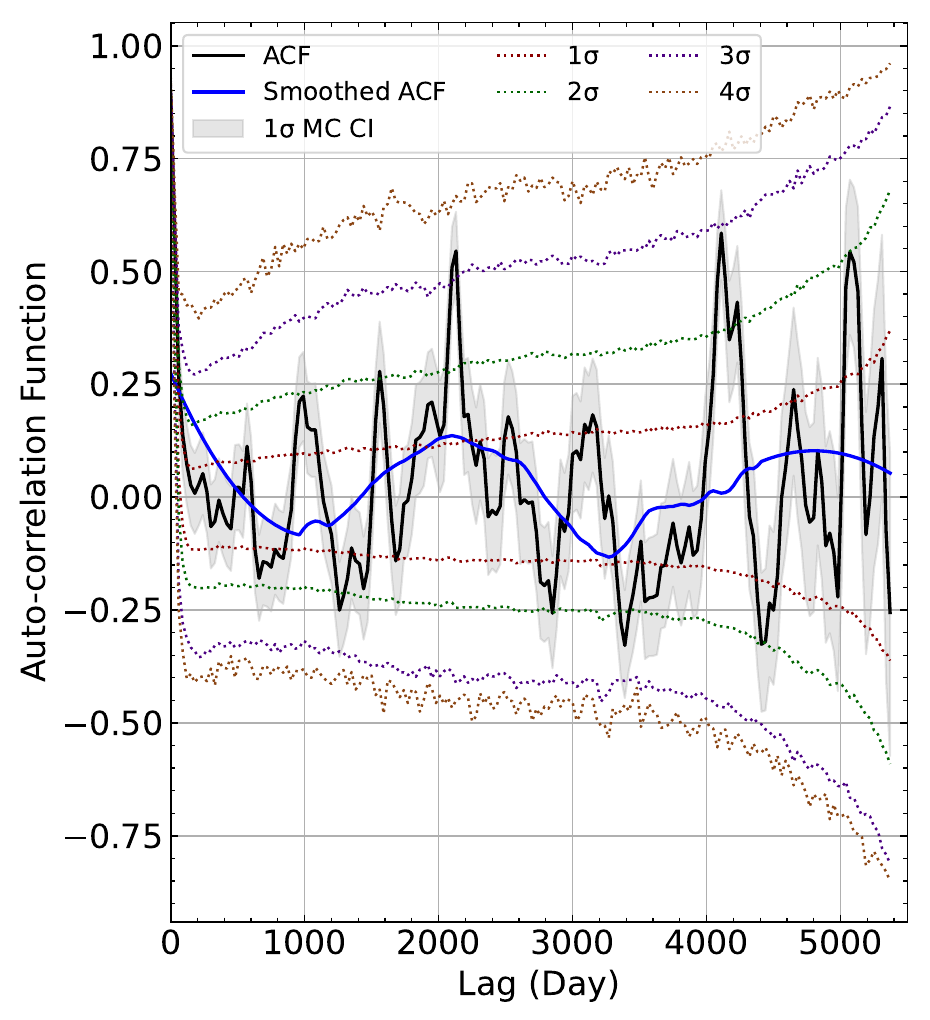}\label{figACF:subfigure2}}
\caption{{Auto-correlation function for PKS 2155-83 (top panel) and PKS 2255-282 (bottom panel) 1-m binned LCs}. The blue lines represent the smoothed correlation, using a Savitzky-Golay filter. The green, red, indigo, and brown lines represent the local 1$\sigma$, 2$\sigma$, 3$\sigma$, and 4$\sigma$ chances of observing the auto-correlation levels indicated by the corresponding lines. These were estimated through simulations of ${10}^5$ LCs, utilizing Emmanoulopoulos' method \citep{Emmanoulopoulos2013}.}
\label{figACF}
\end{figure}

\subsection{Date-Compensated Discrete Fourier Transform}
Another powerful method is the date-compensated discrete Fourier transform (DCDFT), proposed by \cite{Ferraz_Mello_1981}. It  tailors to unevenly spaced data, utilizing the notion of function space projection to realize a Fourier transform. For a given test frequency, the power and amplitude of the DCDFT of unequally spaced data are given by
\begin{equation}
P(w,| x \rangle) = \frac{N[\langle y | y \rangle - \langle 1 | y \rangle^2]}{2S^2}
\end{equation}
and
\begin{equation}
A(w,| x \rangle) = \sqrt{2(\langle y | y \rangle - \langle 1 | y \rangle^2)},
\end{equation}
where $N$ is the number of data points, $y$ is the time simulation function, and $S^2$ is the variance of the time series. The existence of gaps in the data produces spurious peaks in the power spectrum. The CLEANest algorithm can remove spurious peaks, which can be implemented as explained in \cite{Foster_1995}. We used the AAVSO VStar software\footnote{\url{https://www.aavso.org/vstar}} \citep{2012JAVSO..40..852B} to perform the DCDFT and to run CLEANest period analysis refinement algorithm. In what follows, we enclose the DCDFT values obtained via the DCDFT+CLEANest method by parentheses. The timescales quoted were estimated by fitting the power peak to a Gaussian curve, and uncertainty of the signal is the fitting’s half width at half maximum (HWHM).

The PKS 2155-83 1-m binned LC showed two clear peaks at $4.84\pm 0.53$ yr (4.71 yr) and $1.38 \pm 0.05$ yr (1.38 yr). Whereas for the 2-m binned LC, it showed only one such peak at $5.00 \pm 0.69$ yr (4.68 yr), as shown in Fig. \ref{figDCDFT}. In the case of PKS 2255-282, the 1-m binned LC showed three peaks at $1.42\pm0.05$ yr (1.42 yr), $2.79\pm0.17$ yr (2.77 yr), and $5.88\pm0.82$ yr (5.64 yr). Comparable results were obtained for the 2-m binned LC.

\begin{figure}[h]
\centering
{\includegraphics[width=0.49\textwidth]{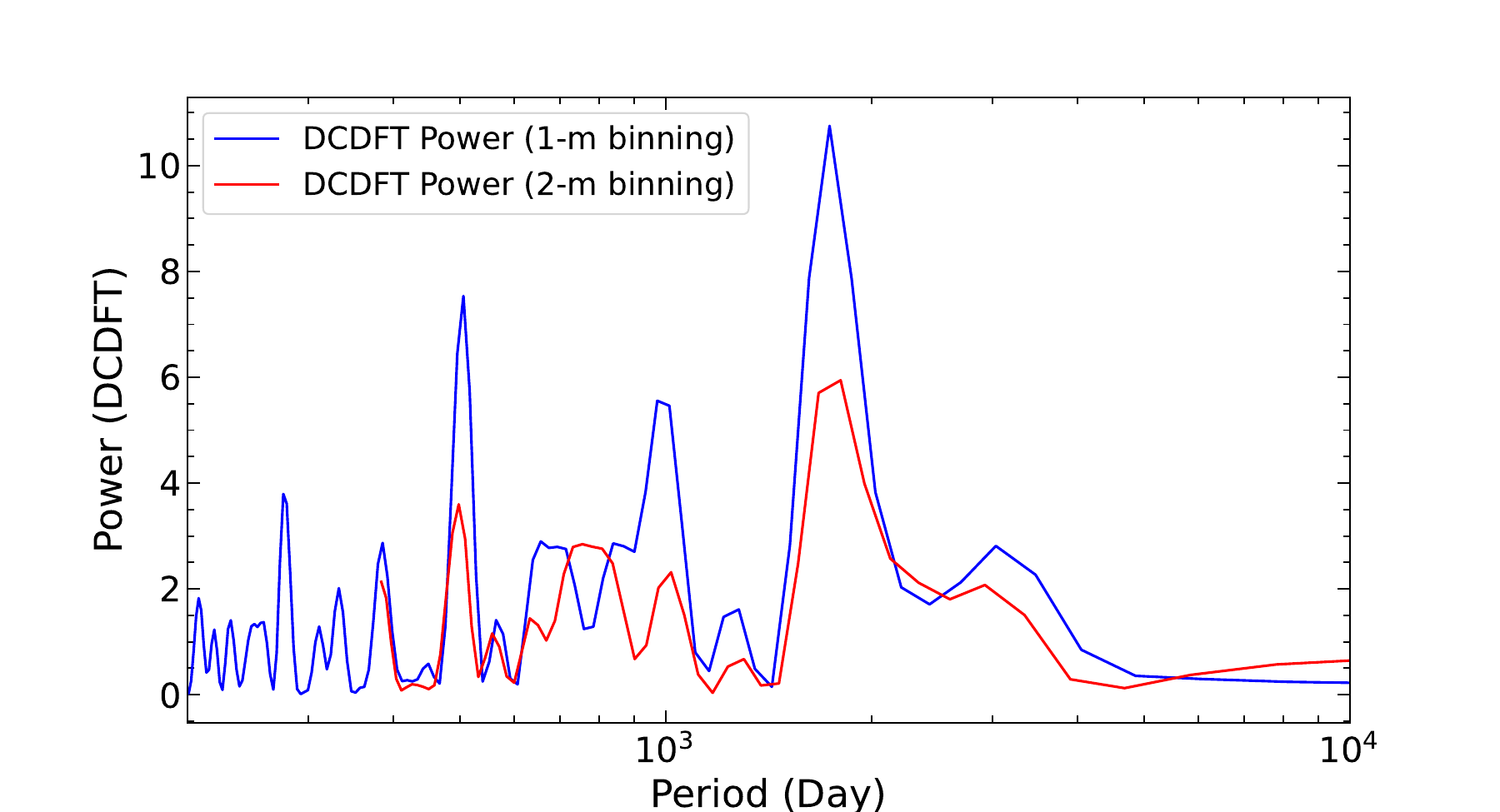}\label{figWWZ:subfigure1}}
{\includegraphics[width=0.49\textwidth]{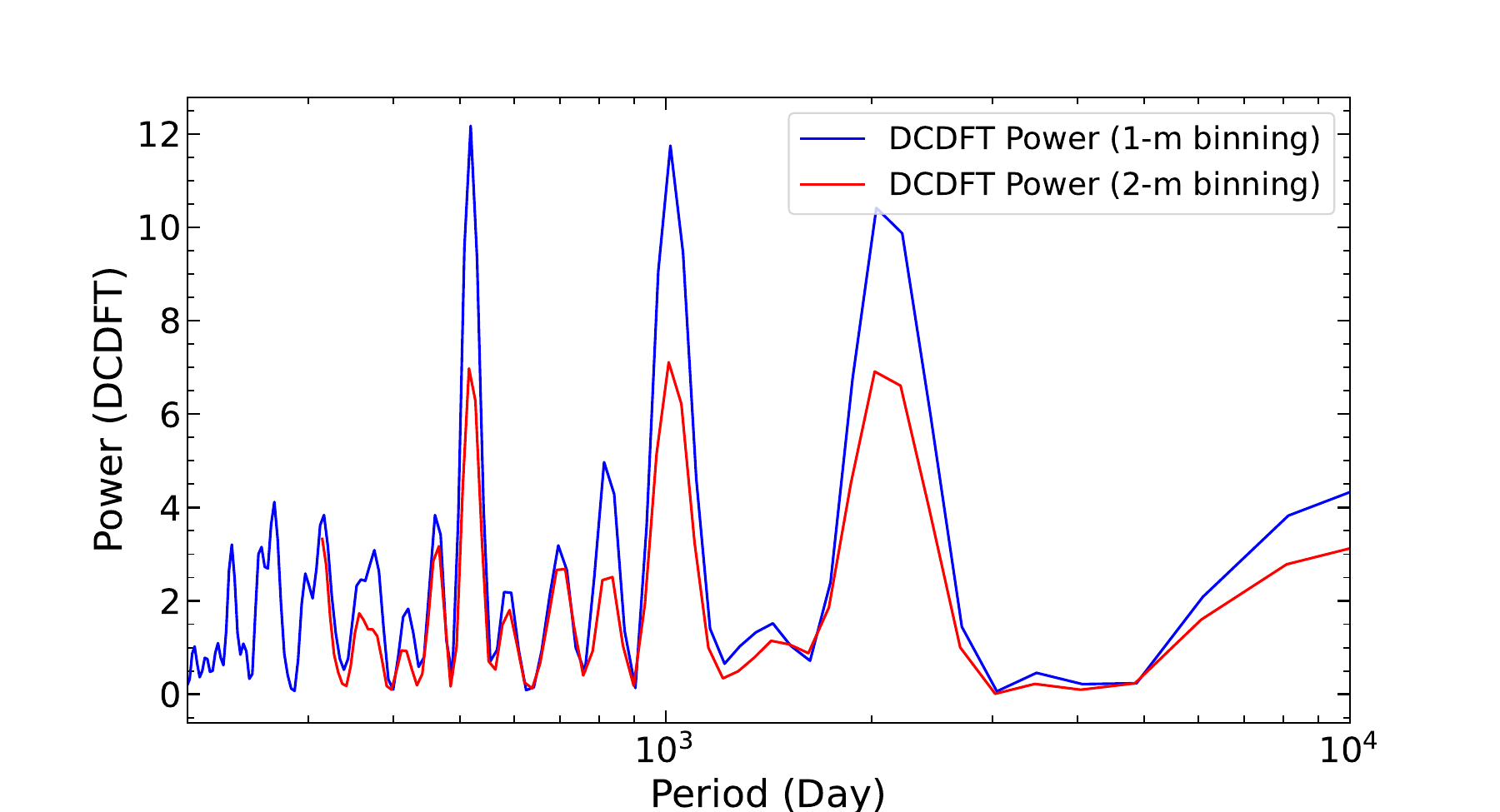}\label{figWWZ:subfigure2}}
\caption{{DCDFT of the 1-m (blue line) and 2-m (red line) binned LCs of PKS 2155-83 (top panel) and PKS 2255-282 (bottom panel).}}
\label{figDCDFT}
\end{figure}

\subsection{The Lomb-Scargle Periodogram}

Further method used to search for quasi-periodicity is the Lomb-Scargle periodogram \citep[LSP;][]{Lomb1976, Scargle1982}, which is a widely used algorithm to establish and characterize periodicity in astronomy, even when the LC has gaps and irregularities. The standard normalized LSP is obtained by fitting the LC to sinusoidal waves of the form $y(t)=A{\mathrm{cos}(\omega t)}+B\mathrm{sin}(\omega t)$. It is defined for a time series $(t_i,y_i)$ as
\[
\begin{aligned}
&P\left(\omega \right)= \\ 
&\frac{1}{2}\left\{\frac{{\left(\sum_i{y_i\ \mathrm{cos}\mathrm{\omega }(t_i-\tau )}\right)}^2}{\sum_i{{\mathrm{cos}}^2\mathrm{\omega }(t_i-\tau )}}+\frac{{\left(\sum_i{y_i\ \mathrm{sin}\mathrm{\omega }(t_i-\tau )}\right)}^2}{\sum_i{{\mathrm{sin}}^2\mathrm{\omega }(t_i-\tau )}}\right\}
\end{aligned}
\]
where, $\tau $ is specified for each frequency to ensure time-shift invariance, such that
\begin{equation}
\mathrm{\tau }\mathrm{=}\frac{1}{2\omega }\mathrm{{tan}^{-1}}\left(\frac{\sum_i{\mathrm{sin}(2\omega t_i)}}{\sum_i{\mathrm{cos}(2\omega t_i)}}\right).
\end{equation}

The generalized Lomb-Scargle periodogram (GLSP) is superior to the standard LSP \citep{Ackermann2015, Prokhorov2017}. Unlike the LSP, the GLSP does not assume that the fitted sine function's mean is the same as the mean of the data. Instead, it accounts for an offset, \textit{c}, to the fitting sinusoidal function, i.e., $\mathrm{y(t)}=A\mathrm{cos(}\omega t\mathrm{)}+B\mathrm{cos(}\omega t\mathrm{)}+c$. For $\gamma$-ray blazars, this term may come from the isotropic diffuse $\gamma$-ray background. In addition, the GLSP takes measurement errors into consideration.

{The GLSP powers of the 1-m binned LC of PKS 2155-83 is shown in Fig. \ref{figGLSP-WWZ}.} In astronomical observations, spurious peaks can spike up due to various contributing factors \citep{VanderPlas2018, Vaughan2016}, e.g., window function aliasing and red noise variability background \citep{Vaughan_2005}. Therefore, the estimated period uncertainty is an essential aspect of reporting the periodogram's results. The false alarm probability (FAP) is one way to quantify peak significance. In the periodograms of PKS 2155-83, the best periods of maximum powers were found to be $T\mathrm{=4.45\pm 0.13\ yr}$ with a FAP of$\mathrm{\ 4.51\times }{10}^{-11}$ and $T\mathrm{=4.42\pm 0.08\ yr}$ with a FAP of $\mathrm{3.64\times }{10}^{-8}$ for the 1-m and 2-m binned LCs, respectively.

Not only the gaps and observation errors influence the periodogram result, but also the bin size has a remarkable impact, particularly on the high-frequency signal. PKS 2155-83 was reported with QPO signal at $T=1.4\pm0.1\ \mathrm{yr}\ (2.8\sigma$) \citep {Peñil2020} in agreement with the present result of $T=1.43\pm0.05\ \mathrm{yr}\ (2.5\sigma$) for the 1-m binned LC. This high-frequency signal disappeared in the 2-m binning LC's periodogram, as the variation details decrease.

To further examine the significance of the power peaks, the ${10}^5$ simulated LCs were used. The period was estimated by fitting the power peak to a Gaussian curve, and its uncertainty is the fitting's HWHM. In this way, a peak was identified with a period of $\mathrm{\ 4.69\pm 0.79\ yr}$ for the 1-m binned LC and $\mathrm{4.55\pm 0.84\ yr\ }$ for the 2-m binned LC at a significance of $(3\sigma )$ for both of them (Table \ref{tableThe estimated periods S1}).

The estimated periods corresponding to the maximum powers in the periodograms of the 1-m and 2-m binned LCs of PKS 2255-282 were $T\mathrm{=6.16\pm 0.27\ yr}$ with a FAP of$\mathrm{\ 1.95\times }{10}^{-8}$ and $T\mathrm{=6.12\pm 0.41\ yr}$ with a FAP of $\mathrm{8.26\times }{10}^{-3}$, respectively. The GLSP of the PKS 2255-282 1-m binned LC showed periods (in years) of $1.43\pm 0.05\ (2.5\sigma )$ and $6.82\pm 2.25\ (2.8\sigma )$ (as shown in Fig. \ref{figGLSP-WWZ}), while the GLSP of the 2-m binned LC showed only the low frequency period at $T\mathrm{=6.73\pm 1.35\ yr}$ with a significance of $2.6\sigma$ (Table \ref{tableThe estimated periods S1}).

\begin{table*}[t]
\centering
\caption{The estimated periods of the $\mathrm{\sim}$15.6 year 1-m and 2-m binned LCs of PKS 2155-83 and PKS 2255-282 with their estimated significance.}
\label{tableThe estimated periods S1}
\begin{tabular}{lccccccc}\hline\hline
Object Name & \multicolumn{1}{c}{Binning} & Power* Pn[zk] & Period T[Pn] & FAP & Chance Prob. & Period T[sim.] & Signif. \\
 & & & [yr] & & & [yr]& \\
\hline
PKS 2155-83\newline & 1-m & 0.38 & $\mathrm{4.5\pm 0.13}$ & $\mathrm{4.5\times }{10}^{-11}$ & $\mathrm{4.8\times }{10}^{-13}$ & $\mathrm{4.69\pm 0.79}$ & $\mathrm{3}\sigma $ \\
& 2-m & 0.51 & $\mathrm{4.4\pm 0.08}$ & $\mathrm{3.6\times }{10}^{-8}$ & $\mathrm{3.3\times }{10}^{-10}$ & $\mathrm{4.55\pm 0.84}$ & $\mathrm{3}\sigma $ \\ \hline
PKS 2255-282 & 1-m & 0.26 & $\mathrm{6.2\pm 0.27}$ & $\mathrm{2.0\times }{10}^{-8}$ & $\mathrm{2.2\times }{10}^{-10}$ & $1.43\pm 0.05$--$6.82\pm 2.25$ & $2.5\sigma $--$2.8\sigma $ \\
& 2-m & 0.21 & $\mathrm{6.1\pm 0.41}$ & $\mathrm{8.3\times }{10}^{-3}$ & $\mathrm{1.8\times }{10}^{-4}$ & $6.73\pm 1.35$ & $2.6\sigma $ \\ \hline
\multicolumn{8}{l}{\parbox{0.9\textwidth}{*The maximum power using the Zechmeister $\&$ Kuerster normalization.}} \\
\end{tabular}
\end{table*}

\begin{figure}[ht]
\centering
{\includegraphics[width=0.49\textwidth]{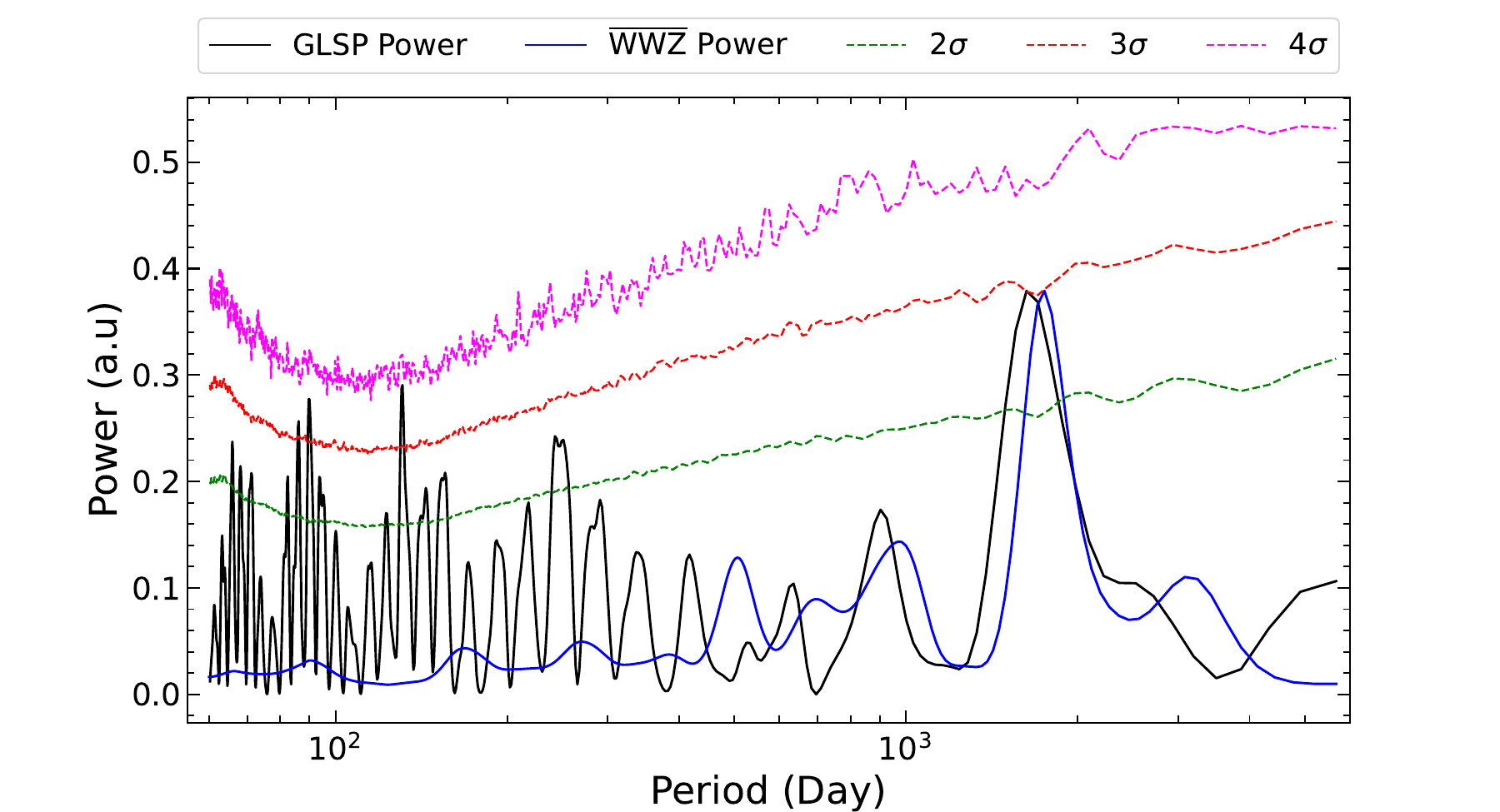}\label{figGLSP-WWZ:subfigure1}}
{\includegraphics[width=0.49\textwidth]{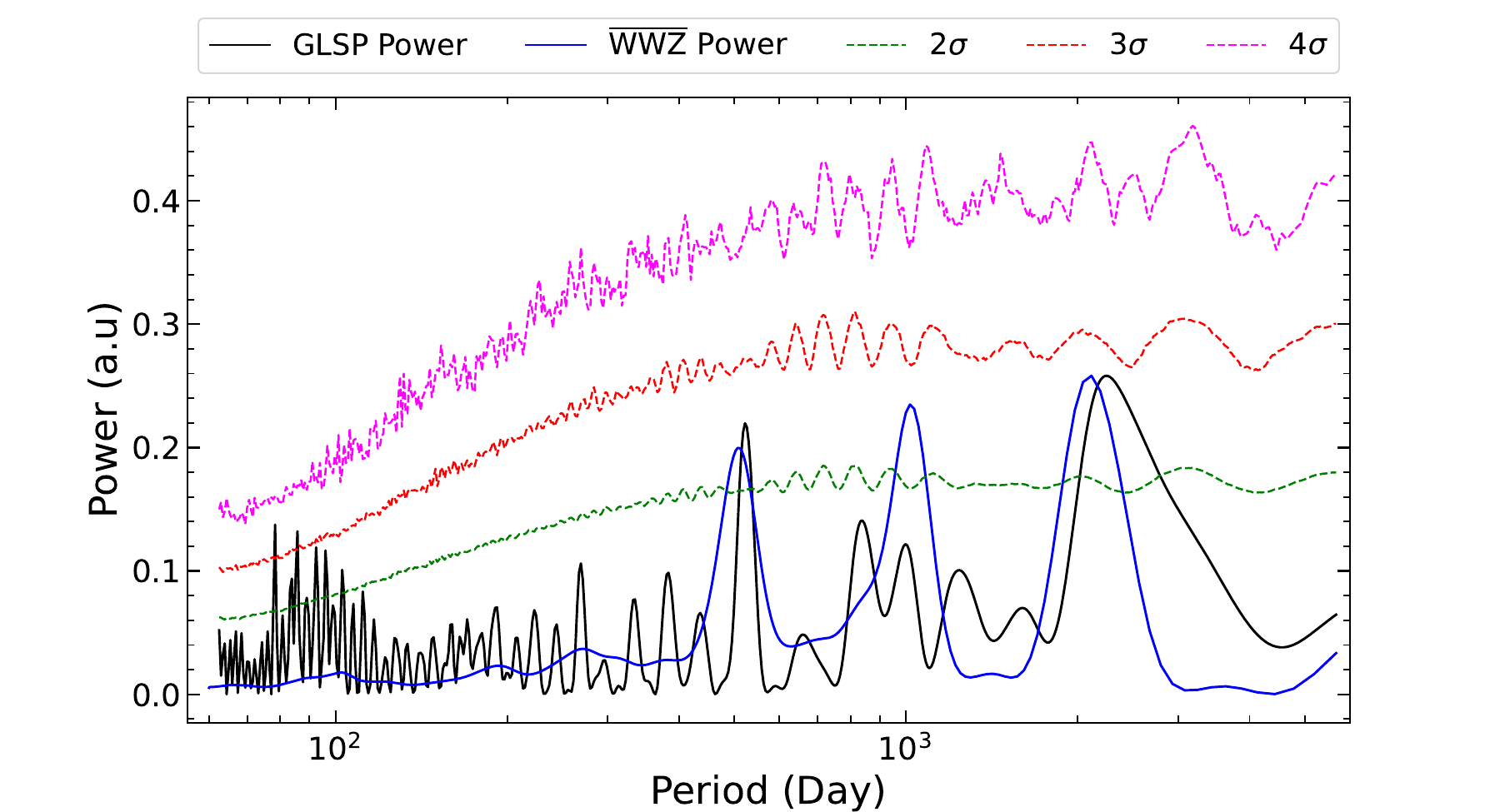}\label{figGLSP-WWZ:subfigure2}}
\caption{{The GLSPs and WWZs of PKS 2155-83 (top panel) and PKS 2255-282 (bottom panel) $\gamma$-ray 1-m binned LCs.} The GLSPs are represented in black lines and the WWZs are represented in blue lines. The green, red, and magenta lines represent the 2$\sigma$, 3$\sigma$, and 4$\sigma$ confidence levels, respectively, of the GLSPs of ${10}^5$ simulated LCs utilizing Emmanoulopoulos' method \citep{Emmanoulopoulos2013}.}
\label{figGLSP-WWZ}
\end{figure}

\subsection{The Weighted Wavelet Z-transform}

An additional efficient method for detecting periodicity associated with LCs of uneven data sampling is the weighted wavelet z-transform \citep[WWZ; ][]{Foster1996}. It is based on a similar notion as the LSP, where the data is fitted by sinusoidal waves. The WWZ can record the possible existence of quasi-periodic variability with a transient nature, where the waves are localized in both frequency and time domains ~\citep{Bhatta2016, Mohan2015, Benkhali2020}.

For the PKS 2155-83 LC, the WWZ gave ({see} Figs. \ref{figGLSP-WWZ} and \ref{figWWZ}) a peak at a period of $4.88\pm 0.61$ yr and $4.93\pm 0.74$ yr for the 1-m and 2-m binned LCs, respectively. The WWZ of PKS 2255-282 gave a period of $5.87\pm 0.85$ yr and $5.81\pm 0.85$ yr for the 1-m and 2-m binned LCs, respectively. In addition, the WWZ of the PKS 2255-282 1-m binned LC displays a notable peak at a period of about 1000 days.

\begin{figure}[ht]
\centering
{\includegraphics[width=0.49\textwidth]{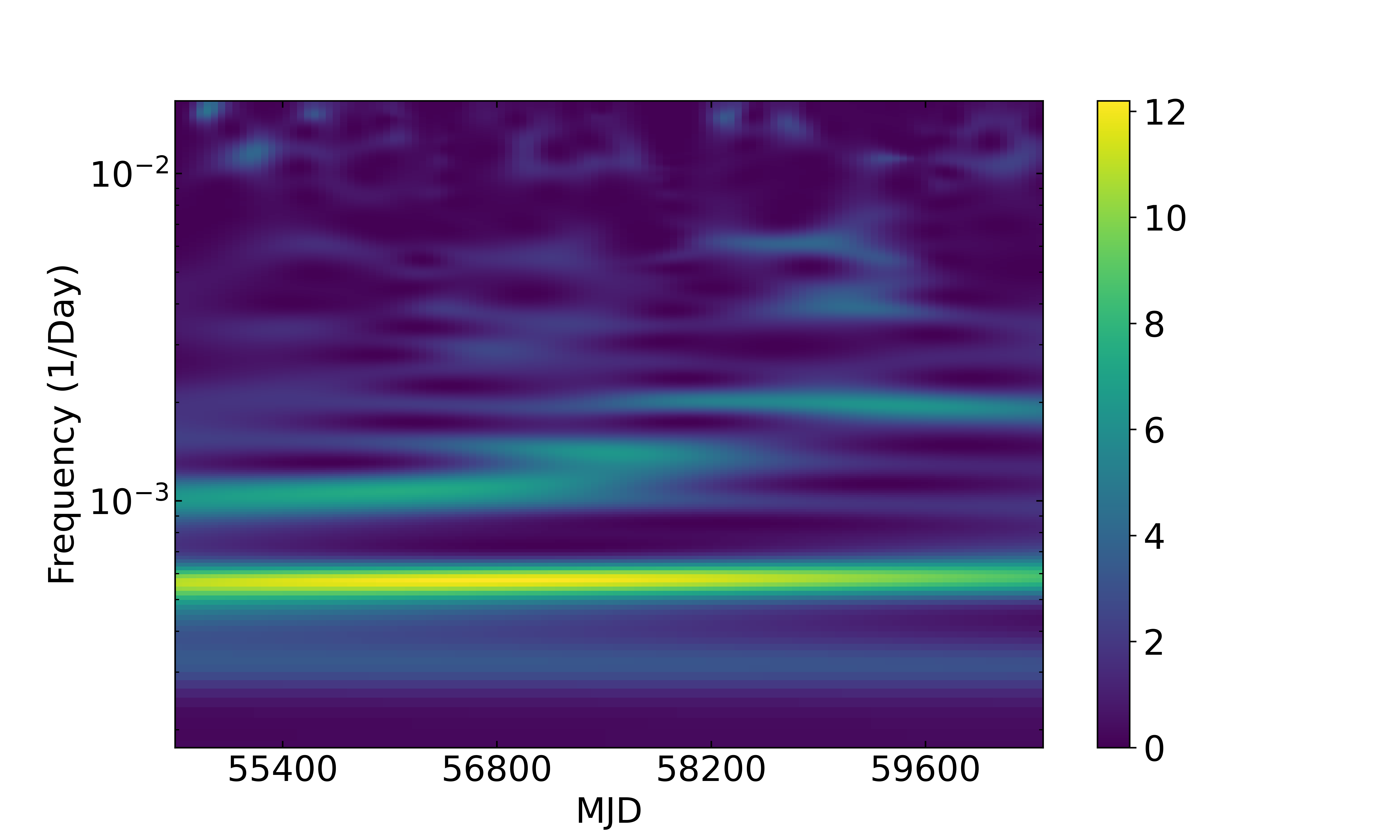}\label{figWWZ:subfigure1}}
{\includegraphics[width=0.49\textwidth]{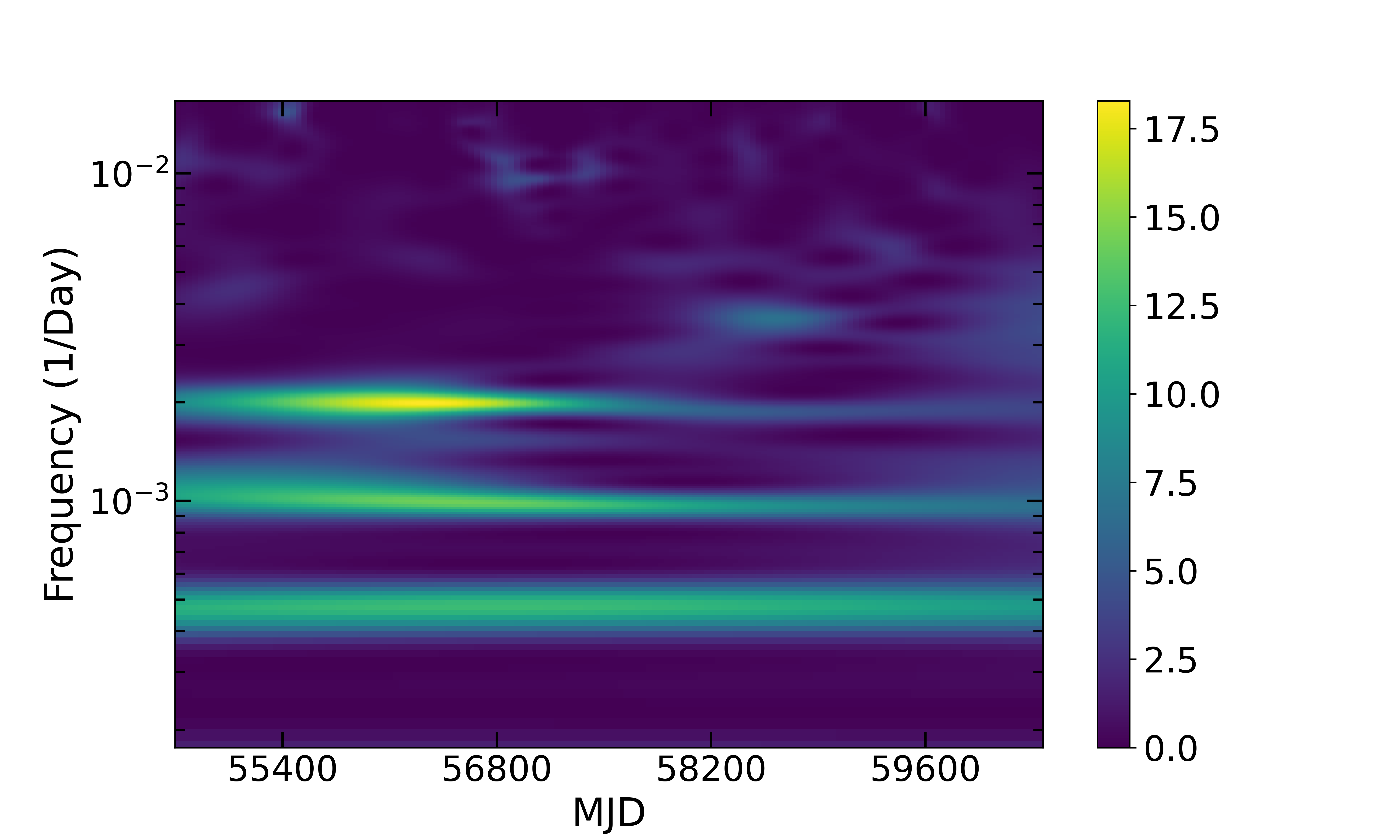}\label{figWWZ:subfigure2}}
\caption{{Two-dimensional contour plots of the WWZ power for the 1-m binned LCs of PKS 2155-83 (top panel) and PKS 2255-282 (bottom panel).}}
\label{figWWZ}
\end{figure}

\subsection{REDFIT}

The LSP and WWZ are strongly affected by red noise at low frequencies, where they place peaks that mimic a real periodicity. The REDFIT method is a suitable tool for detecting periodicity in the dominant red noise LCs of blazars. It simply uses a first-order auto-regressive (AR1) model \citep{Hasselmann1976} to precisely assess the periodogram peaks against stochastic fluctuations \citep{Zhang2021}. The method was coded in Fortran 90 by \cite{Schulz2002}. We used the REDFIT3.8e3 package in the present analysis\footnote{\url{https://www.marum.de/Prof.-Dr.-michael-schulz/Michael-Schulz-Software.html}}.

Compatible with previous estimates, the results for PKS 2155-83 suggest a period of $4.82\pm 0.55$ yr at significance exceeding 99\% and $4.63\pm 0.62$ yr at $2.2\sigma$ (Fig. \ref{figREDFIT}) for the 1-m and 2-m binned LCs, respectively. The results furthermore suggest another peak of $T\mathrm{=1.37\ yr}$ at $2.5\sigma$ and $2\sigma$ for the 1-m and 2-m binned LCs, respectively. For the 1-m and 2-m binned LCs of PKS 2255-282, no significant signal was detected except for a period of about 550 days, as previously reported \citep{Peñil2020}, at significance of $2.3\sigma$ and $2.4\sigma$ for the 1-m and 2-m binned LCs, respectively. It should be noted that the maximum significance provided by REDFIT is limited to $2.5\sigma$.

\begin{figure}[h!]
\centering
{\includegraphics[width=0.49\textwidth]{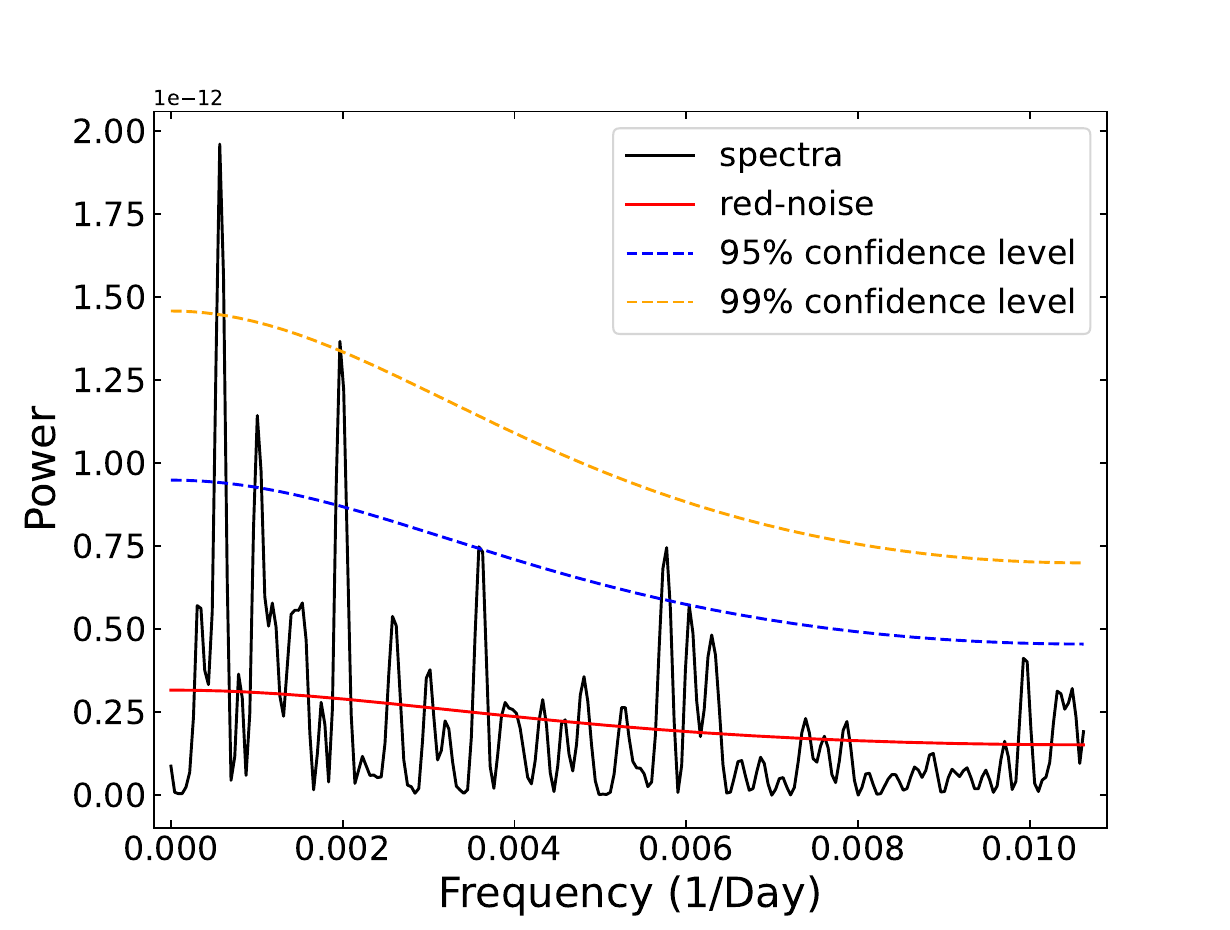}\label{figREDFIT:subfigure1}}
{\includegraphics[width=0.49\textwidth]{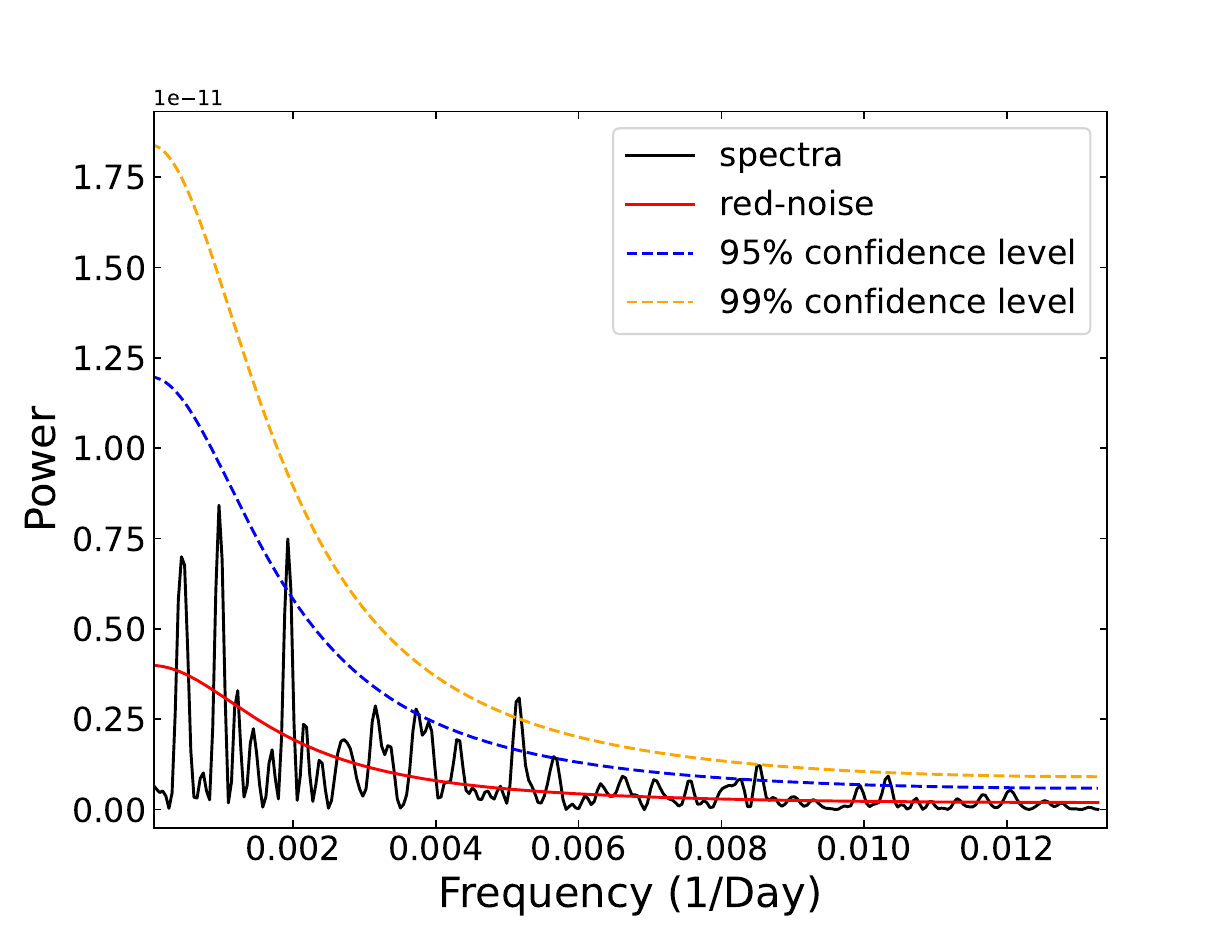}\label{figREDFIT:subfigure2}}
\caption{{REDFIT periodicity analysis for the 1-m binned LCs of PKS 2155-83 (top panel) and PKS 2255-282 (bottom right)}. The bias-corrected power spectrum is represented by the black line. Spectrum of the theoretical red noise (red line), $95\%$ confidence level (blue dashed line), and $99\%$ confidence level (orange dashed line) were estimated by fitting the data with AR1 process.}
\label{figREDFIT}
\end{figure}

\section{Summary and discussion} 
\label{Summary and Discussion}

The growing number of reports regarding the possible existence of quasi-periodicities in the light curves of blazars is interesting; in terms of the information it may embody on the emission processes of these systems, as well as the possibility of detecting binary SMBH systems in the process of merging. A larger sample of candidate objects should include going beyond the well-sampled light curve, where the majority of its time bins have significant observations and only a minority constitute observations with upper limits or gaps.

Here, we studied two such cases of FSRQs from the \textit{Fermi}-LAT 4th catalog. The 1-month and 2-months binned $\gamma$-ray LCs were then generated via the maximum likelihood technique in the energy range $100\ \mathrm{MeV}$--$500\ \mathrm{GeV}$ for about 15.6 years. The selected time bins provided data points with a signal-to-noise ratio above $2\sigma$ for a bin size of 1-month and $3\sigma$ for a bin size of 2-months. The LCs of the two sources showed distinctive behaviors of high and low state alternation. PKS 2155-83 displayed four high states in 2010, 2014, 2019, and 2023, interspersing its otherwise low state. PKS 2255-282 showed three high states in 2009-2013, 2017-2021, and 2023-up to the end of data interrupting its low state.

The probability density function of PKS 2155-83 tended toward log-normality, especially in the case of the 1-month binned LC, indicative of a nonlinear, multiplicative processes  underlying the variability, rather than an additive stochastic process (e.g., associated with many superposed ''mini-jets'' in an AGN engine or simple shot-noise). The probability density function of PKS 2255-282, on the other hand, revealed a comparable degree of fitness with both log-normal and Gaussian distributions.

Various methods were applied to the LCs to assess the possible existence of quasi-periodicity. Namely, the auto-correlation function, {the date-compensated discrete Fourier transform,} the generalized Lomb-Scargle periodogram, the weighted wavelet z-transform, and the REDFIT algorithm. Periodicity peaks were found, using these methods, for PKS 2155-83 in the ranges (in year) $4.45\pm 0.13$--$5.03\pm 0.68$ and $4.42\pm 0.08$--$4.93\pm 0.74$ for 1-month and 2-months binned LCs, respectively. While, for PKS 2255-282, they were found in the ranges of $5.87\pm 0.85$--$6.82\pm 2.25$ and $5.81\pm 0.85$--$6.73\pm 1.35$ for 1 month and 2 months binned LCs, respectively. For  example, the GLSP of PKS 2155-83 showed periods at $4.69\pm 0.79\ \mathrm{yr}$ and at $4.55\pm 0.84\ \mathrm{yr}$ for the 1-month and 2-months binned LCs, respectively. Both with significance $3 \sigma$. The PKS 2255-282 1-month binned LC showed two possible periods at $1.43\pm 0.05\ \mathrm{yr}$ ($2.5 \sigma$) and $6.82\pm 2.25\ \mathrm{yr}$ ($2.8 \sigma$), while the 2-months binned LC showed a period at $6.73\pm 1.35\ \mathrm{yr}$ ($2.6 \sigma$) and no noticeable high frequency signal. {Recently, results of 19 blazars were reported using the first 12 years of data from the \textit{Fermi}-LAT and multiwavelength archival data from radio, infrared, and optical bands \citep{Pe_il_2024}. This study reported no periodic modulations from PKS 2255-282 except of $1.4\pm0.1$ yr (2-3$\sigma$) from the cross-correlation between $\gamma$-rays and the V-band. The disagreement between the suggested periodicity in the present work and the results from the optical observation may argue for the difference between the optical emission region and/or mechanism and the corresponding ones for gamma radiation.}

The existence and origins of QPOs in blazars are still controversial \citep{Sobacchi2016, Sandrinelli2018, Tavani2018}. If confirmed, periodicities may be linked to the process feeding the jet and/or to the relativistic jet itself \citep{Ackermann2015}. They may in general involve scenarios invoking a binary SMBH AGN system (as, for example, discussed in \citealt{Zhou2018}).
Intrinsic origins include possible oscillations associated with instability in the accretion disk or jet formation region \citep{Tchekhovskoy2011}. The characteristic timescales of the pulsational accretion flow instabilities can range from minutes to hours \citep{Honma1992, Tchekhovskoy2012}. 
This is outside the periods suggested in this study.  
Nonetheless, in the case of slow-spinning supermassive black holes, magnetohydrodynamics simulations of magnetically choked accretion flow produce longer frequencies \citep{Tchekhovskoy2012}.

Another possible origin of QPOs in blazars could also be associated with apparent geometrical effects \citep{Rieger2004} e.g., jet precession/helical jet \citep{Caproni2013, Sobacchi2016, Vlahakis1998, Hardee1999, Villata1999, Nakamura2004, Ostorero2004}. In such cases, the observed flux will undergo periodic modulation due to the periodic variation of the Doppler magnification factor \citep{Ackermann2015}. This scenario does not need intrinsic flux modulation and does not induce oscillations in the spectral index.
PKS 2155-83 and PKS 2255-282 showed no flux-photon index correlation along the full time range of observation ($\mathrm{\sim}$ 15.6 yr; with time binning of 1-month and 2-months). Therefore, both objects may favor this second origin. 

Merging supermassive black holes \citep{Begelman1980, Barnes1992} may induce both types of (intrinsic and geometrical) quasi-periodicities. In particular, a SMBH binary system with a milli-pc separation and a total mass of $\sim{10}^8M_{\odot }$ in the early inspiral gravitational-wave driven regime would induce jet precession with timescales of SMBH binary-induced periodicities are ranging from $\mathrm{\sim }$ 1 to $\mathrm{\sim }$ 25 years \citep{Sobacchi2016, Komossa_2014, Rieger_2007}.
Given the redshifts of PKS 2155-83 (z $\mathrm{=1.865}$) and PKS 2255-282 (z ${=0.926}$), reflecting cosmological epochs when merging between galaxies and their embedded black holes were still relatively frequent, the binary black hole scenario may be, in principle, especially relevant. 

We considered a simple model within this general framework \cite{Sobacchi2016}. It assumes a binary system of SMBHs on circular orbits. The direction of the jet, carried by one SMBH, in the center of mass frame is perpendicular to the orbital plane. The jet deviates with an angle $\Delta\alpha$, due to the imprint of the orbital velocity, $v$, of the jet-carrying black hole on the highly relativistic jet. Consequently, the angle $\theta_{\rm obs}$ between the jet and the distant observer oscillates with an amplitude $\Delta\theta_{\rm obs}$, with the same period $T$ of the orbital motion.\footnote{Note that the intrinsic period $T_{\rm int}$ is shorter than the observed period $T_{\rm obs}$ by a factor of $1/(1+z)$.}
This simple model thus incorporates features (particularly, a characteristic timescale) that may be generic to potential blazar quasi-periodic signals originating from binary SMBH systems. 

Based on this model and the periodicities suggested by the GLSP, where $T_{\rm int}$ is 1.64 yr and 3.63 yr, for PKS 2155-83 and PKS 2255-282, respectively, one can estimate the total BH masses. In this context, the total masses (as a function of $\Delta\theta_{\rm obs}$ and mass ratio $q$) are
\begin{equation}
    M=1.4\times {10}^8 {\left(\frac{1+q}{q}\right)}^3 {\left(\frac{\Delta\theta_{\rm obs}}{5^{\circ}}\right)}^3 M_{\odot }
\end{equation}
and
\begin{equation}
    M=3.0\times {10}^8 {\left(\frac{1+q}{q}\right)}^3 {\left(\frac{\Delta\theta_{\rm obs}}{5^{\circ}}\right)}^3 M_{\odot }
\end{equation}
for PKS 2155-83 and PKS 2255-282, respectively.
The separation of the binary is found to be, with $\Delta\theta_{\rm obs}=5^{\circ}$, $R= 0.0035 (1+q)/q\ \mathrm{pc}$ and the SMBHs merging timescale is $T_{\rm GW}= 3.2\times {10}^4 q{\left(q/(1+q)\right)}^3\ \mathrm{yr}$, for PKS 2155-83. A similar exercise gives a separation of the binary $R= 0.0104 (1+q)/q\ \mathrm{pc}$ and SMBHs merging timescale is $T_{\rm GW}= 6.9\times {10}^4 q{\left(q/(1+q)\right)}^3\ \mathrm{yr}$, for PKS 2255-282. Note that the jet has to be assumed to be carried by the secondary SMBH (i.e. $q \gtrsim 1$) to avoid extremely short timescales of orbital decay.

Although the results are quite sensitive to $\Delta\theta_{\rm obs}$, the deflection angle of the highly relativistic jet $\Delta\alpha$, and therefore the amplitude $\Delta\theta_{\rm obs}$, are constrained to a small angle, where $\Delta\alpha\simeq v/c=(q/(1+q)){(GM/R{c}^2)}^{1/2}$ and $\Delta\theta_{\rm obs}=2\Delta\alpha$. In order to have observable consequences on the QPO timescales inferred one needs this angle to be of the order of a few degrees \citep{Sobacchi2016}.

The inferred combined black hole masses are comparable, with reasonable estimation, in both cases considered with independent estimates of the compact central masses estimated in the respective galaxies {(e.g., via dynamical virial equilibrium calculations reported in literature \citealt{Xiong_2014},} where it is found that $\mathrm{log_{10}} (M_{\rm{BH}}/M_{\odot })$ were 9.02 and 9.16 for PKS 2155-83 \citealt{Shaw_2012}, and 8.92 and 9.16 for PKS 2255-282 \citealt{Gu_2001}).
Such supermassive black holes are inferred to be abundant at much higher redshifts. Further searches for high redshift blazars with potential periodicities, in conjunction with gravitational wave signals from SMBH systems from future detectors, such as LISA, may thus serve to test the consistency of the merging black hole scenario as an origin of such signals. This may, in turn, potentially shed light on the SMBH merger rate in the context of standard hierarchical galaxy formation in standard cosmology, ultimately providing a test of the model itself.

\section*{Acknowledgements}

We thank the referee for a careful reading and insightful suggestions that helped improve our manuscript. We also acknowledge the use of Fermitools-conda, DELCgen-Simulating light curves \citep{Connolly2015}, Matplotlib \citep{Hunter2007}, Savgol filter \citep{SciPy-NMeth}, pyZDCF, PyAstronomy \citep{Czesla2019}, NumPy \citep{Harris2020}, AAVSO VStar software \citep{2012JAVSO..40..852B}, REDFIT \citep{Schulz2002}, astroML \citep{Ivezic2014}.

\bibliography{references}{}
\bibliographystyle{aasjournal}



\end{document}